\documentstyle[aps,prl,epsfig]{revtex}
\begin{document}
\draft
\twocolumn[\hsize\textwidth\columnwidth\hsize\csname %
@twocolumnfalse\endcsname
 
\title{$c$-axis conductivity in the normal state of cuprate
superconductors}
\author{ P. Prelov\v sek$^{1,2}$, A. Ram\v sak$^{1,2}$ and I. Sega$^1$}
\address{ $^{1}$ J. Stefan Institute, 1000 Ljubljana, Slovenia }
\address{$^{2}$ Faculty of Mathematics and Physics, University of
Ljubljana, 1000 Ljubljana, Slovenia }
\date{\today}
\maketitle
\begin{abstract}\widetext
The $c$-axis optical conductivity and d.c. resistivity are calculated
within the $t$-$J$ model assuming that the interlayer hopping is
incoherent. Use is made of numerical results for spectral functions
recently obtained with the finite-temperature Lanczos method for
finite two-dimensional systems. In the optimally doped regime we find
an anomalous relaxation rate $\tau_c^{-1} \propto \omega + \xi_c T$ and
$\rho_c(T)\propto\rho_{ab}(T)$, suggesting a common relaxation
mechanism for intra- and interlayer transport. At low doping a
pseudo-gap opening in the density of states appears to be responsible
for a semimetallic-like behavior of $\rho_c(T)$.
\end{abstract}
 ]
\narrowtext

One of the most striking characteristics of cuprate superconductors is
the anisotropy of their structures. The CuO$_2$ planes common to all
high-$T_c$ materials clearly determine most of the normal-state electronic
properties which are quite anomalous but at the same time rather
universal within the whole family of cuprates. The quantity which most
evidently displays the anisotropy of a particular material is the
optical conductivity $\sigma(\omega,T)$ and the corresponding
d.c. resistivity $\rho(T)$ \cite{coop,uchi2}. The ratio
$\zeta(T)=\rho_c/\rho_{ab}$ of the c-axis
resistivity $\rho_c$ to the in-plane resistivity $\rho_{ab}$
for temperatures $T$ just above $T_c$ in optimum-doped materials ranges from
$\zeta\sim 100$  for
YBa$_2$Cu$_3$O$_{7-\delta}$ (YBCO), $\zeta\sim 1000$ for
La$_{2-x}$Sr$_x$CuO$_4$ (LSCO) to $\zeta > 10^5$ in
Bi$_2$Sr$_2$CaCu$_2$O$_8$.

In spite of such large quantitative differences there are common
qualitative features even in the $c$-axis transport, in particular
regarding the variation with doping. (a) Optimum-doped YBCO and
overdoped LSCO show metallic-like $\rho_c(T)\sim a+bT$ for $T>T_c$.
At the same time the optical conductivity $\sigma_c(\omega)$ shows a
fall-off with $\omega$ although far from the usual Fermi-liquid (FL)
Drude-type behavior.  (b) For lower doping the variation is generally
semimetallic-like with $d\rho_c/dT<0$ in the interval $T_c<T<T^*$. The
crossover temperature $T^*$ showing up e.g. as a kink in the in-plane
$\rho_{ab}(T)$, in the Hall constant $R_H(T)$ etc.  \cite{batl},
decreases with doping and merges with $T_c$ at the optimum
doping. Within the low doping regime $\sigma_c(\omega)$ is nearly flat
in a wide frequency regime for higher $T$ while a pseudo gap (PG)
starts to emerge for $\omega<\omega^*$ for lower $T<T_{pg}< T^*$.

The $c$-axis conductivity has attracted the attention of theoreticians
since the first measurements on cuprates \cite{coop}.  In spite of open
challenging questions of anomalous planar properties, there seems to
be an agreement in the conclusion that the $c$-axis conductivity
generally cannot be explained as a coherent interlayer transport
\cite{legg,clar}. The main indication is that the d.c. $\sigma_c$ is
mostly well below the Mott minimum metallic conductivity \cite{mott},
with the possible exception of optimum-doped YBCO and overdoped LSCO
\cite{coop}.  In other words, estimates of the 
mean free path in the $c$-direction
appear much shorter than the interlayer distance $c_0$, i.e. $l_c(T)
\ll c_0$, when one assumes the usual Boltzmann approach for
metals. The absence of coherent $c$-axis conduction is a consequence
of weak interlayer hopping matrix element $t_c$ but also of a strong
intralayer scattering. Based on the concept of dynamical detuning,
Leggett \cite{legg} thus derives the condition for the incoherent
conduction $t_c \ll 1/\tau_a \sim k_{\rm B} T$ taking the
relaxation-rate according to anomalous planar $\rho_{ab}(T)$ and
$\sigma_{ab}(\omega)$. The incoherent conductivity has been in more
detail studied on the example of coupled fermion chains \cite{clar}
representing Luttinger liquids. Still there have been so far no real
attempts to derive more explicitly the $c$-axis conductivity for
models relevant to cuprates.

In this paper we present the theoretical analysis of 
$\sigma_c(\omega)$ within the simplest microscopic model
incorporating both strong electron correlations in the CuO planes
leading to antiferromagnetism (AFM) in undoped materials,
and a weak interlayer coupling. Within each layer we consider the
planar $t$-$J$ model
\begin{equation}
H_l=-t\!\!\sum_{\langle ij\rangle s}(c^\dagger_{ljs}c_{lis}\!+\!
\text{H.c.})\!+\!J\!\sum_{\langle ij\rangle} ({\bf S}_{li}\!\cdot\! {\bf
S}_{lj}\! -\!{1\over 4} n_{li} n_{lj}), \label{tj}
\end{equation}
where $i,j$ refer to planar sites on a square lattice
within the $l$-th layer and
$c_{lis},c^\dagger_{lis}$ represent projected fermion
operators forbidding double occupation of sites.  As appropriate for
cuprates we assume $J=0.3~t$. We allow for the 
hopping between layers,
\begin{equation}
H=\sum_l H_l -t_c\sum_{l i
s}(c^\dagger_{lis}c_{l\!+\!1 is}+\text{H.c.}).
\label{hop}
\end{equation}

The planar $t$-$J$ model is still a challenge for theoreticians,
nevertheless there is a growing consensus that it contains essential
ingredients for an explanation of the anomalous normal-state properties of
cuprates, dominated by effects of strong correlations. The lack of
reliable analytical methods in the latter regime led to numerous
numerical studies \cite{dagorev}. Using the novel finite-temperature
Lanczos method (FTLM) \cite{jplanc} it has been shown that several
thermodynamic quantities as well response functions match well as the 
experimental data on cuprates. Most relevant for the present study are
the results on the planar conductivity $\sigma_{ab}(\omega)$ \cite{jpcon} and
the single-particle spectral functions (SF) $A({\bf k},\omega)$ \cite{jpspec}.

In the limit of weak interlayer hopping $t_c \ll t$ it is
straightforward to evaluate the dynamical $c$-axis conductivity
$\sigma_c(\omega)$ using the linear response theory. Expressing the
$c$-axis current correlation function $\langle j_c(t) j_c \rangle$, with
$j_c = e_0 t_c c_0 \sum_{lis}(ic^\dagger_{lis}c_{l\!+\!1 is}+\text{H.c.})$, 
in terms of the planar SF $A({\bf k},\omega)$, we arrive at
\begin{eqnarray}
\sigma_c(\omega)&=& {\sigma_c^0\over \omega}\int
{d\omega^\prime}
[f(\omega^\prime)-f(\omega^\prime+\omega)] \nonumber \\
&&\times {4 \pi t^2\over N}\sum_{{\bf k}} { A}({\bf k},\omega^\prime)
{A}({\bf k},\omega^\prime+\omega), \label{sigc}
\end{eqnarray}
where $\sigma_c^0= e_0^2 t_c^2 c_0/\hbar a_0^2 t^2$ is a
characteristic $c$-axis conductivity scale, $N$ is the number of sites and
$f(\omega)=[1+\exp(\omega/k_{\rm B}T)]^{-1}$.  The approximation
assumes the independent electron propagation in each layer, and is
justified for $t_c\ll t$.  Note that the interlayer-hopping term,
Eq.~(\ref{hop}), {\it conserves} the QP momentum ${\bf k}$, and in this respect
the treatment is analogous to the one of the transverse transport in
weakly coupled Luttinger chains \cite{clar}. This could be compared
with a related problem of interlayer hopping with random matrix elements
$(t_c)_{li}$, where only the planar density of states (DOS) ${{\cal
N}}(\omega)= 2/ N \sum_{{\bf k}} { A}({\bf k},\omega-\mu)$ enters,
with $\mu$ denoting the chemical potential. The corresponding
expression for the $\bf k$-{\it nonconserving} 
$\sigma_c^n(\omega)$ is obtained by the replacement
\begin{eqnarray}
\sigma_c^n(\omega)&=&{\sigma_c^0\over \omega}
\int d\omega^\prime
[f(\omega^\prime)-f(\omega^\prime+\omega)]  \nonumber \\
&&\times{\pi t^2}{ {\cal N}}(\mu+\omega^\prime){{\cal
N}}(\mu+\omega^\prime+\omega), \label{sign}
\end{eqnarray}
appearing e.g. in disordered systems \cite{mott}, together with a
substitution for $t_c$ in the definition of $\sigma_c^0$ with some average
$\bar{t_c}$.  In cuprates both
alternatives have a justification, since the disorder introduced by
dopands residing between layers modifies the hopping elements. Hence
one can expect that the actual conductivity is a linear combination of
$\sigma_c$ and $\sigma_c^n$.

The knowledge of planar $A({\bf k},\omega)$ and ${\cal N}(\mu+\omega)$
should thus suffice for the evaluation of $\sigma_c(\omega)$. In
cuprates, however, SF are available with sufficient resolution only in
the hole part $\omega<0$ (in principle) via the angle-resolved
photoemission (PES) \cite{shen} and the DOS via the angle integrated
PES \cite{ino}.  Within the $t$-$J$ model (and the closely related
Hubbard model) $A({\bf k},\omega)$ and ${\cal N}(\omega)$ have been
studied mostly numerically \cite{dagorev}, applying the exact
diagonalization (ED) of small systems \cite{hors} and the quantum
Monte Carlo method \cite{bulu}. Results reveal at intermediate doping a large
Fermi surface (consistent with the Luttinger theorem) and a
quasiparticle (QP) dispersion $\epsilon_{\bf k}$ similar to but
reduced in bandwidth relative to the free tight-binding electrons. 
Only recently, via
the FTLM \cite{jplanc}, a reliable evaluation of the corresponding
self energy $\Sigma({\bf k},\omega)$ was possible, thus allowing for a
study of low-$T$ QP properties.  The SF and $\sigma_{ab}(\omega)$ both
reveal in this regime the anomalous behavior consistent with the
marginal Fermi liquid (MFL) concept \cite{varm}, i.e. the effective
transport relaxation rate as well as the QP damping appear to follow 
$1/\tau_{ab}(\omega)\propto \Sigma''({\bf k},\omega) \propto |\omega|
+\xi_{ab} T$.

In this paper we use numerical results for $A({\bf k},\omega)$ and
${\cal N}(\omega)$, as obtained using FTLM on systems with $N=16,18,
20$ sites for various hole dopings $c_h=N_h/N$. SF are available for
the whole range of $N_h$ for $N=16$, $N_h\leq 2$ for $N=18$, and
$N_h\leq1$ for $N= 20$ \cite{jpspec,jplanc}.  One should keep in mind
the restriction that FTLM results become dominated by finite-size
effects for $T<T_{fs}$. $T_{fs}$ is clearly size and doping
dependent. The lowest $T$ can be reached in the intermediate regime
$c_h\sim 0.2$ where $T_{fs} \sim 0.1~t$, while $T_{fs}$ increases both
for lower and higher doping. To make contact with experiments, we note
that $t\sim 0.4$~eV, i.e. our numerical results apply to $T\agt 450$~K, while
for  lower $T$ they can give some qualitative indications.

To set the frame for further discussion let us first present in Fig.~1
the DOS ${\cal N}(\mu+\omega)$ for the two lowest nonzero doping
levels $c_h=1/20$ and $c_h=2/18$ for several $T\leq J$. The DOS at 
larger doping $c_h \agt 0.15$ (not shown here)
\cite{jpspec} is $T$-independent and featureless at $\omega \sim 0$,
as expected for a Fermi liquid away from van Hove singularities.
At low doping the DOS in Fig.~1 is also structurelesss for $T>J$, but here 
this reflects the incoherent character of QP due to large damping.
Upon lowering $T$ a gradual transfer of weight in the underdoped samples
from above $\mu$ to $\omega<0$ is observed, leading to the formation
of a PG at $\omega\sim 0$. The deepening is more pronounced for lowest
dopings, e.g.  $c_h=1/20$ in Fig.~1(a), where the PG energy scale is
well below $t$ and appears to be $\Delta_{pg}\sim J$, reflecting the
onset of the short-range AFM order.
\begin{figure}[htb]
\begin{center}\leavevmode\epsfxsize=78mm\epsfbox{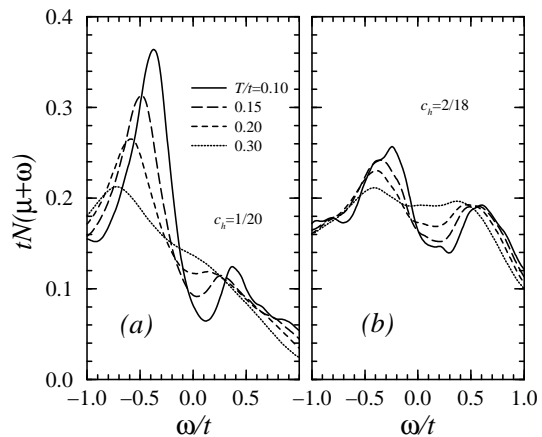}\end{center}
\caption{The DOS ${\cal N}(\mu+\omega)$ at various $T\leq J$ for hole
concentrations: (a) $c_h=1/20$ and (b) $c_h=2/18$. }
\end{figure}
The above development of the DOS as well as the behavior of SF and
corresponding (anomalous) self energies should determine the behavior
of optical conductivity $\sigma_c(\omega)$.  In Fig.~2 we show
$\sigma_c(\omega)$ for two dopings $c_h=3/16$ and $c_h=1/20$
for $T\leq J$, as calculated from Eq.~(\ref{sigc}).
At intermediate doping $c_h=3/16$ $\sigma_c(\omega)$ in Fig.~2 exhibits a 
\begin{figure}[htb]
\begin{center}\leavevmode\epsfxsize=70mm\epsfbox{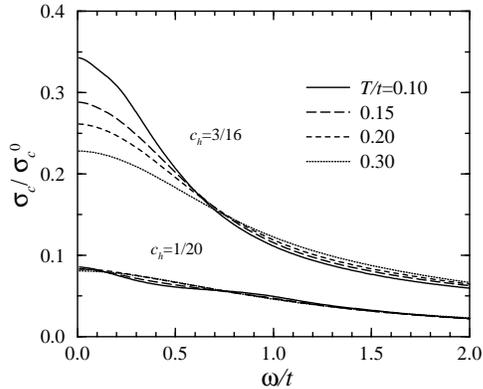}\end{center}
\caption{$\sigma_c(\omega)$ at several $T$ for dopings:
$c_h=3/16$ and $c_h=1/20$.}
\end{figure}

\noindent
central peak, sharpening at lower $T$.  This can be
explained with the properties of $A({\bf k},\omega)$ in this regime,
consistent with a large Fermi surface and a QP dispersion $\epsilon_{\bf k}$
corresponding  to a tight-binding band. The QP damping is however
large and MFL-like $\Sigma''({\bf k},\omega) \propto |\omega|
+\xi_{ab} T$, but only weakly ${\bf k}$-dependent. Inserting the
latter into Eq.~(\ref{sigc}) it is easy to see that the resulting
$\sigma_c(\omega)$ also follows the MFL-behavior, i.e. one can
represent it in the generalized Drude form with an effective
relaxation rate $1/\tau_c\propto |\omega| + \xi_c T$. The vanishing rate
for $\omega,T \to 0$ is clearly the consequence of the ${\bf
k}$-conservation in the interlayer hopping. On the other hand,
Eq.~(\ref{sign}) would yield a qualitatively different result, i.e. an
almost flat and $T$-independent $\sigma_c^n(\omega)$ due to the 
structureless DOS ${\cal N}(\omega)$ for $\omega \sim 0$.

The corresponding results for low-doping $c_h=1/20$ are also shown in
Fig.~2. In contrast, we find rather weakly $\omega$- and $T$-dependent
$\sigma_c(\omega)$, both for the $\bf k$-conserved and the
$\bf k$-nonconserved approximation. Quantitative agreement of both
approaches indicates that in this regime we are dealing with a strongly
reduced effective dispersion $\epsilon_{{\bf k}} \propto J$ but still
large QP damping, hence weakly ${\bf k}$-dependent $A({\bf k},\omega)
\sim {\cal N}(\mu+\omega)$, at least for $T>T_{fs}$. It seems puzzling
how rather constant $\sigma_c^n(\omega)$ in Fig.~2 for
$c_h=1/20$ can be compatible with a well pronounced PG at $\omega \sim
0$ in the DOS (see Fig.~1) at low $T<T^*$.  These facts are however
reconciled by observing that the Fermi functions in
Eqs.~(\ref{sigc},\ref{sign}) have an effective width $\sim 4T$, hence
down to reachable $T_{fs}$ they smear out both the QP dispersion as
well as the $\Delta_{pg}$. Nevertheless, there is some signature of a PG
opening on a scale $\omega\sim J$ in $\sigma_c(\omega)$ for lowest
$T$, in qualitative agreement with experimental data \cite{uchi}.

In Fig.~3 we present the d.c. resistivity
$\rho_c(T)=1/\sigma_c(\omega=0,T)$ for  a
wider  range of $c_h$. In the regime of intermediate and even higher doping
$c_h \geq 3/16$  $\rho_c(T)$ is metallic-like for all $T$.  In
particular from the previous arguments relating the $c$-axis transport to
the MFL-like planar relaxation one expects that $\rho_c(T) \propto T$ for
$T<J$. From Fig.~3 we realize that the latter behavior is really
restricted to $T<0.2~t$.
It should be stressed again that in this regime weak interlayer disorder as
implied by Eq.~(\ref{sigc}) is {\it essential} for the
resulting linear-in-$T$ behavior of $\rho_c(T)$, whereas $\rho_c(T)$ in the
case of $\bf k$-nonconserving interlayer hopping (not shown in Fig.~3) yields
an almost  $T$-independent $\rho_c(T)$.
\begin{figure}[htb]
\begin{center}\leavevmode\epsfxsize=70mm\epsfbox{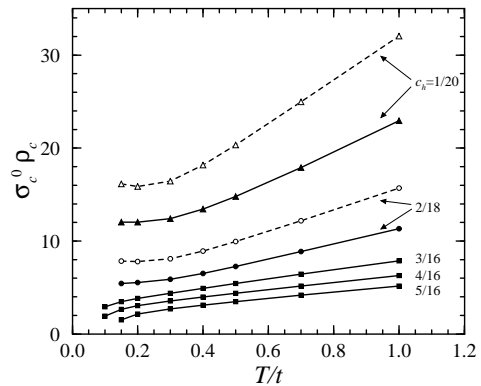}\end{center}
\caption{ $\rho_c(T)$ vs. $T$ for various $c_h$. Solid and
da\-shed lines represent results within the $\bf k$-conser\-ving and 
$\bf k$-nonconserving approximation, respectively.} 
\end{figure}

For the lowest $c_h=2/18$ and $c_h=1/20$ $\rho_c$ and $\rho_c^n$
yield the same qualitative behavior,
$\rho_c^n(T)$ being systematically higher. 
While $\rho_c(T)$ is increasing for $T>J,t$
as generally expected for incoherent transport, the relevant question is its
behavior for $T<J$. In Fig.~3 we notice that in this regime
$d\rho_c/dT \sim 0$, signalling the onset of a nonmetallic-like
behavior in agreement with experimental results for cuprates in the
underdoped region \cite{naka}.

When we comment on the relation of our results with measurements on
cuprates, the most natural application is to LSCO having the  simplest
layered structure among cuprates. The presence of several nonequidistant
layers per unit cell, and possibly chains as well, in other materials
complicate the interpretation of the $c$-axis
transport considerably. In Fig.~4 we present the resistivity ratio
$\zeta(T)=\rho_c/\rho_{ab}$ as a function of $T$ for different doping
levels, where results for $\rho_{ab}$ are taken from previous FTLM
studies \cite{jpcon}. In the saturation regime $T\agt J$ we 
set $\zeta\sim 150$ as
obtained from  experimental data for LSCO e.g. by Nakamura and Uchida
\cite{naka} and Kao {\it et al.} \cite{kao} from which we can estimate
$t_c$.  Taking the standard value for $c_0/a_0=3.5$ in LSCO
\cite{coop} we get $t_c\sim 0.03t$, or $t_c\sim 12$~meV.
 
A distinctive feature is that $\zeta(T)$ is almost independent of
$c_h$ and $T$, but for the lowest $T<J$ presented. This indicates that planar
resistivity and c-axis resistivity are related, at least for $T>T_{fs}$,
pointing to a common current relaxation mechanism in all
directions. In fact, if we calculate within the same decoupling scheme
the planar conductivity $\sigma^{\rm dec}_{ab}(\omega)$ and the related
$\rho^{\rm dec}_{ab}(T)$, the actual $\sigma_{ab}(\omega)$ is reproduced
remarkably well, as can be
concluded from $\zeta_{ab}(T)=\rho^{\rm dec}_{ab}/\rho_{ab}$ in Fig.~4. 
For $T\sim J$ we detect the onset of a different behavior,
although the ratio is less affected than $\rho_{ab}(T)$ and
$\rho_c(T)$ separately. It is however indicative that the increase of
$\zeta(T)$ is most pronounced for lowest $c_h$.  Our results for
$\zeta(T)$ are in a remarkable qualitative agreement with experimental
data~\cite{kao}.
\begin{figure}[htb]
\begin{center}\leavevmode\epsfxsize=64mm\epsfbox{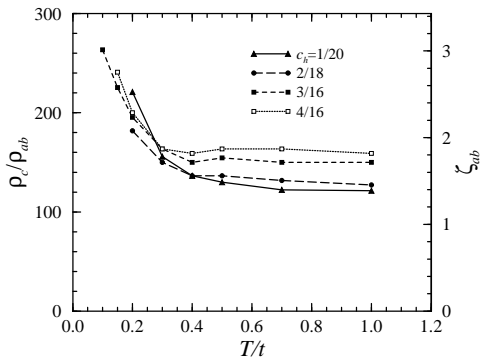}\end{center}
\caption{$\rho_c/\rho_{ab}$ vs. $T$ for various $c_h$ (left scale) and
$\zeta_{ab}(T)=\rho^{\rm dec}_{ab}/\rho_{ab}$
(right scale).}
\end{figure}

In conclusion, let us briefly summarize our main results on the
$c$-axis conductivity.  In our analysis $\sigma_c(\omega,T)$ is obtained
from (numerically) known SF and DOS for the planar $t$-$J$ model
supplemented with a small (doping independent) 
interlayer hopping $t_c$ leading to
the incoherent $c$-axis transport.  A qualitative agreement with experiments
in LSCO, e.g. for the anisotropy $\zeta(T)$, gives support to such
a minimum model. Our results also confirm the general
experimental observation that $\rho_c(T)$ changes from a metallic one
to a semimetallic one $d\rho_c(T)/dT \alt 0$ by decreasing doping. It
should be however noted, that in the overdoped regime 
LSCO shows already indications of an enhanced coherent-like $c$-axis
transport which is beyond our analysis assuming $t_c \ll t$.

Let us finally point on open questions, mainly related to the onset of
a PG in $\sigma_c(\omega)$ and even semiconductor-like $d\rho_c(T)/dT
\ll 0$ as found experimentally in underdoped cuprates \cite{uchi2}.
In our analysis the indication for such a development comes from a
PG in the DOS, which at low doping starts to emerge for $T<T^*\alt J$,
and is compatible with recent PES experiments \cite{ino}.  Clearly,
this phenomenon is related to the onset of short-range AFM spin
correlations, persisting up to the optimum doping. It is also
evident from our results that this does not lead directly to a well
pronounced PG in $\sigma_c(\omega)$ at the same $T$. Eventually this
can happen only for $T <T_{pg}< T^*$, which in our systems is
hardly reachable due to $T_{fs}\sim 0.2\,t$ at low doping.  One simple
explanation can be given in terms of the Fermi-function broadening
$\sim 4T$. Another source of suppression of the PG in calculated
$\sigma_c(\omega)$ is the shift of the PES leading edge towards $\mu$
(see Fig.~1), thus compensating in part for the deepening of the PG
in the DOS at lowest $T$. This results in a much weaker doping
dependence of $\rho_c$ down to $T\sim T_{fs}$, not
inconsistent with experimental data \cite{naka} which show
semiconductor-like upturn only at lower $T$. Still it remains the
subject of future studies to clarify whether the small PG scale in underdoped
cuprates is directly related to the $T^*$ scale or is of different origin.

\end{document}